\documentclass[preprint2]{aastex631}

\usepackage{graphicx} % Required for inserting images
\usepackage{xfrac}
\usepackage{xcolor}
\usepackage{amsmath,amssymb,amsfonts}
\usepackage{latexsym}
\usepackage{epstopdf} 
\usepackage{wasysym}
\usepackage{float}
\usepackage{array,multirow}

\begin{document}

\title{It's always the quiet ones: Single Degenerate Double Detonation Type Ia Supernova from Quiescent Helium Accretion}
% \title{Single Degenerate Double Detonation Type Ia Supernova from Helium Accretion}

\author[0000-0002-1361-9115]{Amir Michaelis}
\affiliation{Department of Physics Technion — Israel Institute of Technology Haifa 3200003, Israel}

\author[0000-0002-0023-0485]{Yael Hillman}
\affiliation{Department of Physics, Azrieli College of Engineering,
Ramat Beit Hakerem, POB 3566, Jerusalem 91035, Israel}
\affiliation{Department of Physics Technion — Israel Institute of Technology Haifa 3200003, Israel}

\author[0000-0002-5004-199X]{Hagai B. Perets}
\affiliation{Department of Physics Technion — Israel Institute of Technology Haifa 3200003, Israel}
\affiliation{Astrophysics Research Center of the Open University (ARCO), The Open University of Israel, P.O Box 808, Ra'anana 4353701, Israel}

%\author{Amir Michaelis}
%\date{July 2024}

%\begin{document}

%\maketitle

\begin{abstract}
    We investigate a sub-Chandrasekhar mass double detonation pathway for Type Ia supernovae arising from single degenerate helium accreting carbon-oxygen white dwarfs. Building on our previous one dimensional study of recurrent helium novae \citep{hillman2025}, we evolve a $0.7~\rm{M_\odot}$ white dwarf through steady accretion at $10^{-8}~\rm{M_\odot~yr^{-1}}$ until it reaches $1.1~\rm{M_\odot}$, yielding realistic, time evolved helium rich profiles. These profiles are mapped into FLASH simulations, incorporating nuclear burning for helium and carbon–oxygen detonation, in multi-dimensional hydrodynamic runs. A localized, modest temperature perturbation near the base of the helium shell robustly triggers an outward helium-shell detonation. The ensuing inward propagating shock converges in the carbon–oxygen core, igniting a secondary detonation that unbinds the star. We obtain a $^{56}$Ni yield of $\simeq 0.64 \rm{M}_\odot$, an intermediate-mass element (Si-Ca) mass of $\simeq 0.41 \rm{M}_\odot$, and maximum ejecta velocities approaching $\sim 22,000~\rm{km~s^{-1}}$ - values consistent with normal Type Ia supernovae. Our results demonstrate that recurrent helium accretors-typically quiescent over long timescales-can evolve under subtle, ``quiet" conditions to trigger robust double detonations, supporting their role as viable progenitors of sub-Chandrasekhar mass Type Ia supernovae.
\end{abstract}


\section{Introduction}\label{sec:intro}

Type Ia supernovae (SNe Ia) are indispensable tools for cosmology \citep{brout2019,galbany2023,palicio2024} and major contributors to galactic iron-peak enrichment \citep{shikui2010,soma2014,cavichia2024,souradeep2025}, yet decades of study have not yielded a consensus on their progenitors \citep{howell2011,maoz2014,hoeflich2017,jha2019,liu2023,soker2024,soker2025,Ruiter2025}. Classical single-degenerate (SD) models \citep{Whelan1973,Nomoto1982I} invoke a carbon–oxygen white dwarf (CO WD) accreting from a non-degenerate companion until reaching the Chandrasekhar mass ($\sim\!1.4\,\mathrm{M_\odot}$), where ignition proceeds via a deflagration-to-detonation transition \citep{Khokhlov1991,Plewa2004,Gamezo2004,Gamezo2005} or a gravitationally confined detonation in multi-dimensional simulations \citep{Calder2004,Meakin2009,Jordan2008,Jordan2012}. However, deep late-time nebular spectra of nearby SNe Ia (e.g., SN~2011fe) show no $\mathrm{H}\alpha$ or He\,\textsc{i} emission down to $\lesssim\!10^{-4}\,\mathrm{M_\odot}$ of stripped companion material, and pre-explosion imaging together with early-time UV limits rule out a bright red-giant, placing stringent constraints on classic SD channels \citep{Li2011,Nugent2011}. Observations of SN~2022xkq and SN~2023bee—showing fast declines and/or early flux excesses—pose additional challenges to current progenitor and explosion models \citep{Pearson2024,Wang2024}.

An alternative double-degenerate (DD) pathway involves the merger of two WDs via gravitational-wave–driven inspiral. If the combined mass exceeds $\sim\!1.4\,\mathrm{M_\odot}$, central carbon ignition may occur (``classical" DD; \citealt{Webbink1984,Iben1984}), but off-center burning can instead form an O–Ne WD rather than a supernova \citep{Saio1998,Shen2012}. More recently, ``dynamically driven double-degenerate double detonations" ($\mathrm{D}^6$; \citealt{Shen2010,Pakmor2013,Boos2024,Pollin2025}) and head-on collisions \citep{Raskin2009,Kushnir2013,Glanz2025} have been shown to produce prompt, sub-Chandrasekhar detonations in one or both WDs. Both SD and DD channels predict delay-time distributions (DTD) $\propto t^{-1}$ over $0.1$–$10$~Gyr \citep{Friedmann2018,Friedmann2021}, consistent with observed SN~Ia rates in old stellar populations, yet neither channel cleanly satisfies all spectroscopic and rate constraints.

Recent work has also highlighted the role of hybrid CO–He white dwarfs in the DD landscape. In these systems, one or both WDs possess a CO core surrounded by a thick helium mantle. During the merger, the less massive hybrid can tidally disrupt and deposit a substantial He layer onto the primary, creating conditions for an edge-lit detonation at sub-Chandrasekhar mass \citep{Perets2019,Pakmor2021}. Population-synthesis studies suggest that hybrid (He–CO) WDs are common in close binaries and that a large fraction of double-WD mergers involve at least one hybrid. In some models, CO+hybrid mergers contribute at levels comparable to CO+CO mergers to the overall SN~Ia rate and DTD, though assumptions about common-envelope efficiency and metallicity can shift the balance \citep{Nelemans2001,Toonen2017,Liu2017}.

In parallel, \citet{soker2025} has proposed a double-degenerate merger-to-explosion-delay (DD-MED) scenario, in which two CO WDs merge and explode after a $\sim\!1$–$2$~yr relaxation phase, potentially explaining SN~Ia interactions with CO-rich circumstellar material at $\gtrsim\!50$~days post-explosion. This delayed detonation within a merger remnant again highlights that “quiet,” non-traditional channels may hide within the SN~Ia population.

Against this backdrop, the sub-Chandrasekhar ``double-detonation" scenario has emerged as a robust alternative: a CO WD accretes a thin He shell that detonates first, sending a shock into the core that triggers a secondary carbon–oxygen detonation \citep{Nomoto1982II}. Multi-dimensional studies demonstrate that even shells as low as $\sim\!0.03\,\mathrm{M_\odot}$ can focus a strong inward shock and robustly detonate the core—yielding normal-luminosity SNe~Ia without fine tuning \citep{Pakmor2022}. Synthetic light curves predict early UV/optical ``bumps" from shell ashes, and high-velocity Ca\,\textsc{ii}/Ti\,\textsc{ii} absorption features in the first days—features that are now being sought in high-cadence surveys \citep{Yao2019,Piro2025}.

A critical uncertainty in these models has been the physical origin and composition of the He shell. 
One scenario that in certain cases may allow for a helium shell to gradually accumulate on the surface of a WD, is as the product of nova eruptions. In this case, the system would be a cataclysmic variable (CV) where mass in transferred through the inner Lagrangian point (L1) \cite[]{Shara1981, Prikov1995, Yaron2005, Shara2018, Hachisu2010}, or a symbiotic system, for which the wind from a red giant (RG) or an asymptotic giant branch star (AGB) is either gravitationally captured by the WD via the Bondi-Hoyle-Littleton (BHL) mechanism, \cite[]{Mikolajewska1992, Mikolajewska2008, Mikolajewska2010, Hillman2021a, Vathachira2024, Vathachira2025b} or if the binary orbit is close enough, via wind focusing through L1, i.e., wind Roche-lobe overflow (WRLOF) \cite[]{Abate2013, Saladino2019, Vathachira2025a}. Regardless of the mass transfer mechanism, the result is hydrogen-rich material being accumulated on the surface of the degenerate WD, which eventually ignites in a runaway process and produces a nova eruption \cite[e.g.,][]{Gallagher1978, Starrfield1987, Warner1995}. Models have shown that if the hydrogen (H) is accreted at a high enough rate, the eruption will not eject all the mass that has been accreted, but will leave behind a He-rich residue \cite[]{Kato1999, Kato2004, Hillman2016}. This helium may accumulate over many H-novae, until eventually reaching its own critical mass that triggers a He-nova \cite[]{Jose1993, Idan2013, Newsham2013}. Based on this, we \cite{Hillman2015} followed the results of the one-dimensional nova-simulation code accreting hydrogen at a high rate, and used the average accumulation rate of He-rich material as the input for direct helium accretion. Our results showed periodic helium novae that, after a long adaptation period, became less degenerate and led to the growth of the CO core.

Acknowledging that helium could accumulates on the surface of a WD via the bi-product of nova eruptions, or via direct accumulation from a He-star (or a He-WD), leading to a wide possibility of accumulation rates, \citet{hillman2025} followed helium-accretion on CO WDs ($0.65$–$1.0\,\mathrm{M_\odot}$) at different rates $\dot{M}\sim10^{-5}-10^{-10}\,\mathrm{M_\odot\,yr^{-1}}$, showing that while the high rates produce recurrent novae, for the regime $\lesssim10^{-8}\rm M_\odot yr^{-1}$  helium need not ignite promptly; the system may avoid nova eruptions and continue to accrete helium until a thermonuclear runaway (potentially a SN~Ia or an undetermined transient, UT). These models produce helium shells of order a few $\times\!10^{-2}\,\mathrm{M_\odot}$ up to a few $\times\!10^{-1}\,\mathrm{M_\odot}$ that naturally accumulate before runaway at sub-Chandrasekhar masses. These time-evolved profiles—complete with realistic thermal and compositional gradients—offer physically motivated initial conditions for multi-dimensional detonation simulations, in contrast to ad hoc shell prescriptions.

Here, we unite detailed He-nova evolution with multi-dimensional hydrodynamics by mapping an initial $0.7\,\mathrm{M_\odot}$ WD that quiescently accretes He-rich material at a rate of $\dot{M}=10^{-8}\,\mathrm{M_\odot\,yr^{-1}}$ until reaching $1.1\,\mathrm{M_\odot}$. We entered the WD profile into FLASH simulations \citep{Fryxell2000}, incorporating 19-element nuclear burning for both helium and carbon–oxygen. By seeding a modest temperature perturbation at a random point on the shell–core interface, we capture the double-detonation mechanism—the outward He-shell detonation, inward shock convergence, and subsequent core detonation. We then quantify explosion energetics, nucleosynthesis yields, and ejecta kinematics, providing a unified, physically motivated demonstration that quiescent helium accretors can robustly produce normal-luminosity SNe~Ia in a double-detonation single-degenerate scenario. Section~2 describes our 1D and FLASH setups; Section~3 presents our results; Section~4 discusses implications for progenitor demographics and observations; and Section~5 summarizes our conclusions.

\subsection{Double-detonations with thick vs.\ thin helium shells}
Early sub-Chandrasekhar studies showed that detonating a sufficiently massive
helium (He) shell can drive a converging shock into the C/O core and ignite a
secondary detonation \citep[e.g.,][]{Livne1990,WoosleyWeaver1994}.
These ``thick-shell'' models (typically $M_{\rm He}\!\gtrsim\!0.1~M_\odot$)
also highlighted a tension: extensive He-shell burning at high velocity seeds
outer iron-group and intermediate-mass material (notably Ti, Cr, Ca) that can
produce strong line blanketing and redder colors than normal SNe~Ia
\citep{WoosleyWeaver1994}. Motivated by this, the field pivoted to ``thin-shell''
realizations where $M_{\rm He}\!\lesssim\!0.05~M_\odot$ suffices to trigger the
core while minimizing high-velocity shell ash
\citep{Fink2010,Kromer2010,WoosleyKasen2011}.

Our setup deliberately revisits the historical thick-shell regime, but with
physically time-evolved, quiescent He envelopes as initial conditions for the
explosion phase. This lets us test whether shock focusing and core ignition remain
robust under modern numerics while reassessing how outer IME/IGE ash maps onto
observables (radiative transfer is deferred to a follow-up paper). The hydrodynamic
outcome here, a shell detonation that triggers a core detonation, is consistent
with multi-D demonstrations of ignition robustness across shell masses
\citep{Livne1990,Fink2010}.

\subsection{Relation to prior simulations}
Multi-D work since the 1990s finds that once a lateral He detonation wraps the
star, the inward shock can reach core-ignition conditions over broad parameters
\citep{Livne1990,Fink2010}. Where models diverge is in the {\em observable}
imprint of the shell ash. Thin, modestly polluted shells
($M_{\rm He}\!\lesssim\!0.05~M_\odot$) tend to yield bluer colors and spectra
closer to normal SNe~Ia \citep{Kromer2010,WoosleyKasen2011}, whereas thicker
shells enhance Ca/Ti/Cr at high velocity and can redden the early-time SED
\citep{WoosleyKasen2011}. Systematic grids connecting shell mass/composition to
light-curve and spectral behavior \citep{Polin2019} further indicate that the
``massive-shell'' regime better matches peculiar, strongly line-blanketed cases
(e.g., SN~2018byg) than the bulk of normal SNe~Ia.

Our fiducial model lies on the ``thicker-shell'' side of these mappings, but with
a stratification and small residual He mass that may mitigate the most extreme
color effects (see Section~\ref{sec:caveats}). Here we emphasize the dynamical
robustness and nucleosynthetic budgets and quantify the outer-shell IME/IGE yields
against the literature ranges; detailed radiative transfer is beyond this paper's scope.

\section{Methods}
\label{sec:methods}

\begin{figure*}
  \centering
  \includegraphics[width=0.65\textwidth]{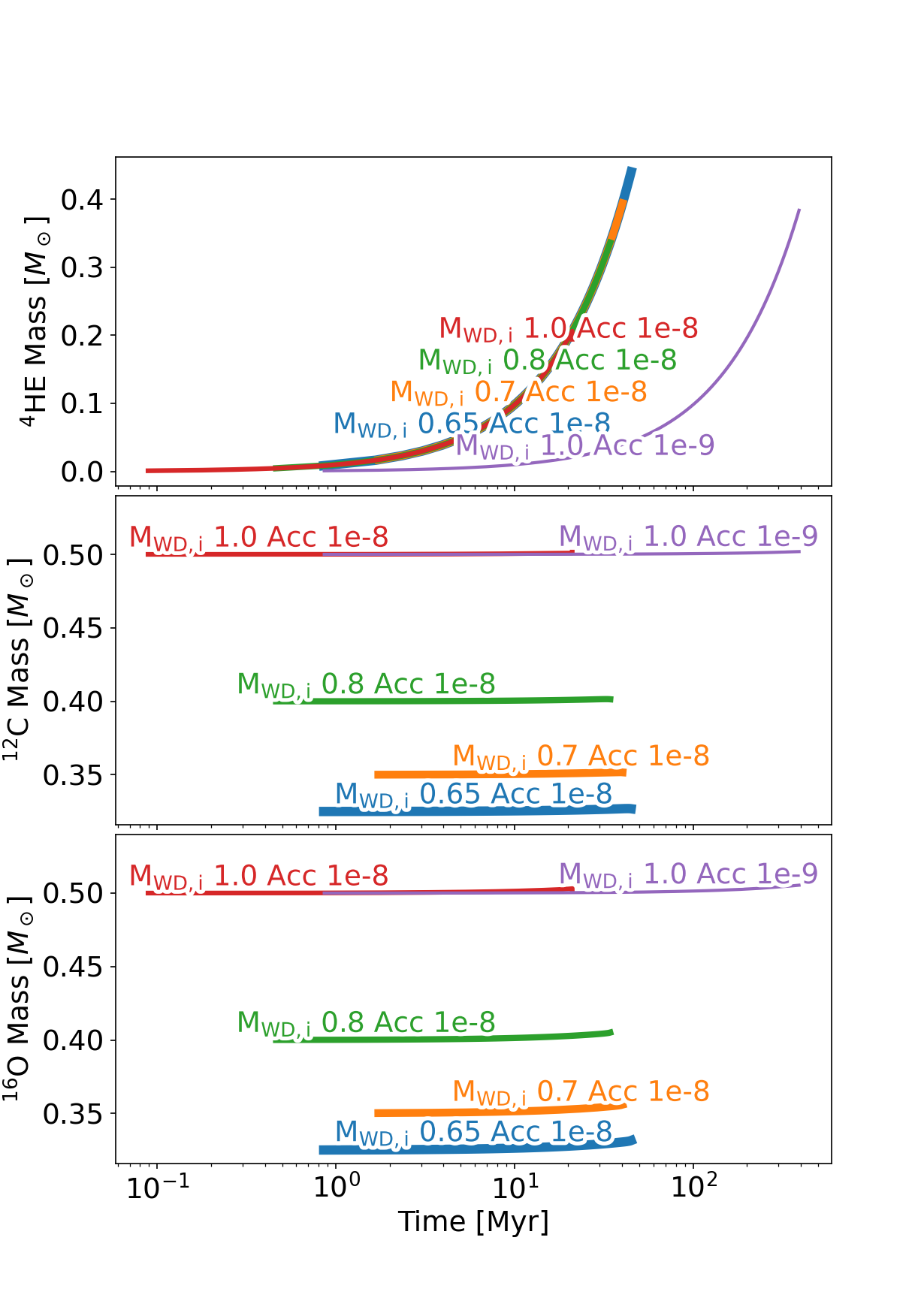}
  \caption{\textbf{Long-term envelope and core evolution in quiescent accretion models.}
  Helium mass (top) and bulk core carbon/oxygen masses (middle/bottom) versus time for several $(M_{\rm WD,i},\dot M)$ tracks in \citet{hillman2025}. Our fiducial model (orange) starts at $M_{\rm WD,i}=0.7\,\mathrm{M_\odot}$ with $\dot M=10^{-8}\,\mathrm{M_\odot\,yr^{-1}}$ and grows to $1.1\,\mathrm{M_\odot}$. These models experienced uninterrupted prolonged helium accretion until indicating a SN ignition.}
  \label{fig:acc_ele}
\end{figure*}

We begin from the quiescent helium-accretion model \#23 of \citet{hillman2025} (see there, Tables~1 and 3). A $0.7\,\mathrm{M_\odot}$ carbon–oxygen (CO) white dwarf (WD) accretes helium at $\dot{M}=10^{-8}\,\mathrm{M_\odot\,yr^{-1}}$ and grows to $1.1\,\mathrm{M_\odot}$. The one–dimensional (1D) hydrodynamic Lagrangian nova code used by \citet{hillman2025} follows consecutive eruptions from helium accretion. Originally developed for hydrogen–rich accretion \citep{Prikov1995,Yaron2005,Epelstain2007,Hillman2015}, the code was adapted for helium accretion as in \citet{Hillman2016} to quantify how much helium a WD can retain prior to a thermonuclear runaway (TNR). We extract a snapshot immediately before explosive burning (i.e., before Si–group burning becomes numerically unresolved in 1D), yielding hydrostatic radial profiles $\{\rho(r),T(r),X_i(r)\}$ of a helium–rich envelope over a CO core as functions of enclosed mass and radius. Figure~\ref{fig:acc_ele} summarizes the long–term envelope growth and core composition evolution for several $(M_{\rm WD,i},\dot M)$ tracks from \citet{hillman2025}; our fiducial track is $M_{\rm WD,i}=0.7\,\mathrm{M_\odot}$, $\dot M=10^{-8}\,\mathrm{M_\odot\,yr^{-1}}$.

The explosive phase is modeled with \textsc{flash}~4.7 \citep{Fryxell2000}, solving the compressible Euler equations in three dimensions with the unsplit solver \citep{Lee2006,Lee2009,Lee2013} and self–gravity via the multipole Poisson solver. Microphysics includes the Helmholtz electron–positron Fermi–Dirac equation of state (degenerate/relativistic electrons and positrons, ideal ions with Coulomb corrections, and radiation; \citet{Aparicio1998,Timmes1999,Timmes2000}), and nuclear energy generation with the approximate 19–isotope $\alpha$–chain network \citep{Weaver1978}, which captures the dominant energy–releasing flows for helium burning and CO detonation.

Profiles are remapped from the Lagrangian grid to the \textsc{flash} Eulerian mesh using a mass–conservative interpolation that preserves zone–integrated mass, momentum, and internal energy. We then perform a brief hydrostatic relaxation with nuclear reactions disabled, self–gravity enabled, and a velocity–proportional damping term that decays exponentially over a few dynamical times. After relaxation, the model is hydrodynamically stable, and the central density and temperature remain within $<\!1\%$ of their mapped values.

Adaptive mesh refinement (AMR) is applied \emph{only} on density and temperature gradients so as to track the detonation front and the core–shell interface cleanly. The finest cell size is $\simeq 7~\mathrm{km}$, ensuring enough fine cells across the numerically broadened fronts in both the helium shell and the upper CO core.

\begin{figure*}
  \centering
  \includegraphics[width=0.65\textwidth]{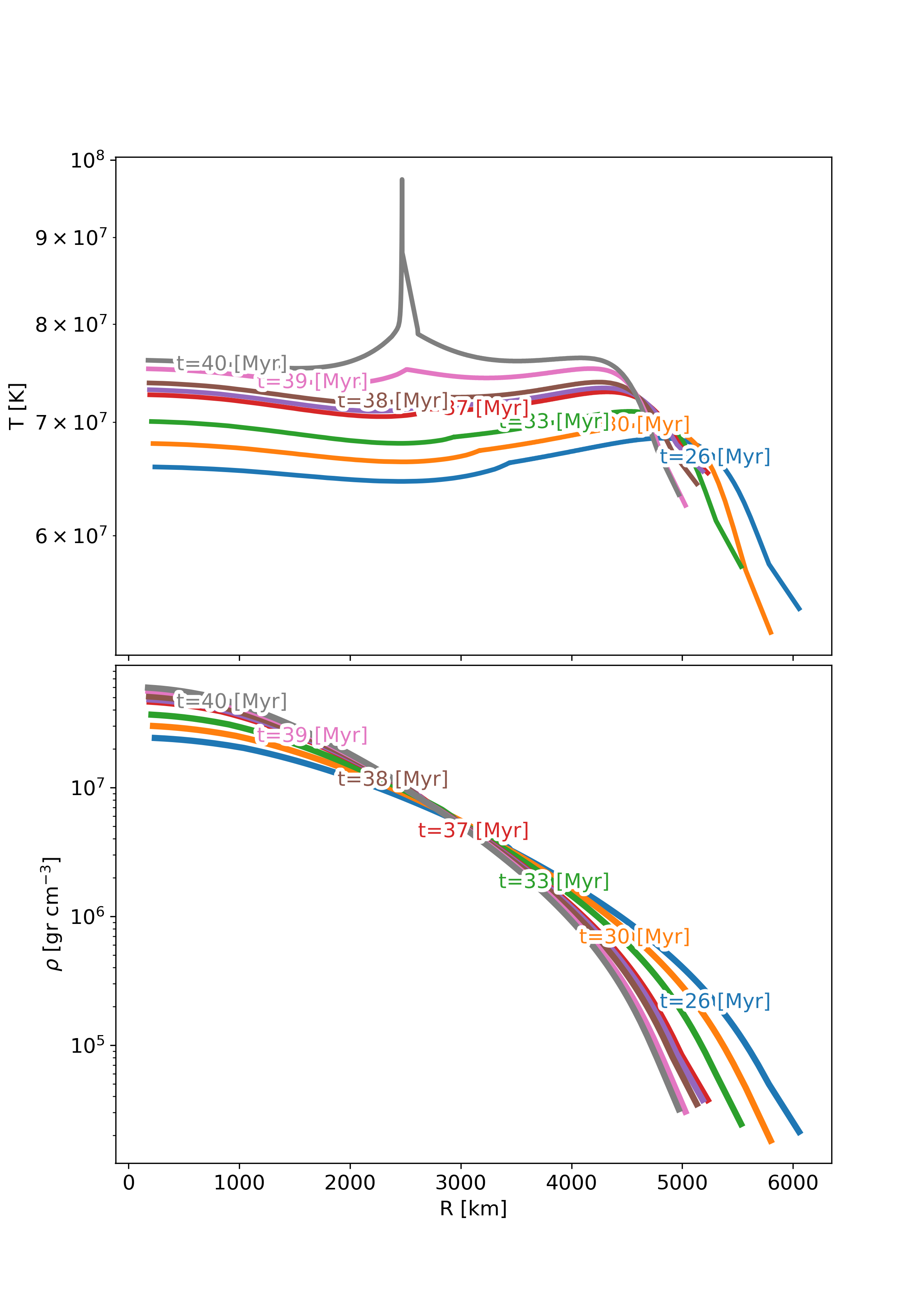}
  \caption{\textbf{Pre-ignition structure during uninterrupted helium accretion.}
  Temperature (top) and density (bottom) as functions of radius at several epochs in our fiducial 1D Lagrangian model.
  The narrow temperature spike at $r\simeq2.5\times10^3$~km marks the naturally emerging base-of-shell peak reached a few timesteps before the 1D calculation becomes numerically unresolved ($T_{\max}\approx3\times10^8$~K).
  In the multi-D \textsc{flash} simulations we do not map this peak to an entire shell; instead, we represent it as a localized off-axis Gaussian hotspot to avoid artificially overheating a shell while preserving the physical trigger.}
  \label{fig:wd_prof}
\end{figure*}

To seed ignition with a minimal perturbation, we introduce an \emph{off–axis} Gaussian temperature hotspot centered at $(r,\theta,\phi)=(2470~\mathrm{km},\,60^\circ,\,30^\circ)$ with characteristic radius $\sigma_r=200~\mathrm{km}$ and peak temperature $T_{\max}=3\times 10^8~\mathrm{K}$. In the 1D Lagrangian models, the local base temperature naturally approaches this perturbed value prior to runaway; however, mapping such a temperature to an extended shell in 3D would artificially inflate the heated mass. We therefore apply the perturbation to a single point (resolved region) only in order to avoid overheating the shell while still capturing a physically plausible hotspot trigger. The resulting pre–ignition thermal and density structures are shown in Figure~\ref{fig:wd_prof}.

For nucleosynthesis diagnostics, we seed $10^6$ Lagrangian tracer particles uniformly in a box enclosing the WD and retain those with sufficiently high pre–detonation density to undergo significant burning, yielding a working set of $\sim 4\times 10^5$ tracers. For each tracer we record $(t,\rho,T,\{X_i\})$ at the reaction substep cadence and post–process with a 127–isotope network \citep{Paxton2015} to obtain elemental/isotopic yields and composition–versus–velocity structures. %Volume–integrated cell–center abundances provide a cross–check and agree with tracer–integrated yields at the few–percent level.

\section{Results}
\label{sec:results}

\begin{figure*}
  \centering
 %\fbox
 { \includegraphics[trim={8.0cm 4.0cm 6.0cm 5.0cm}, clip, width=0.98\textwidth]{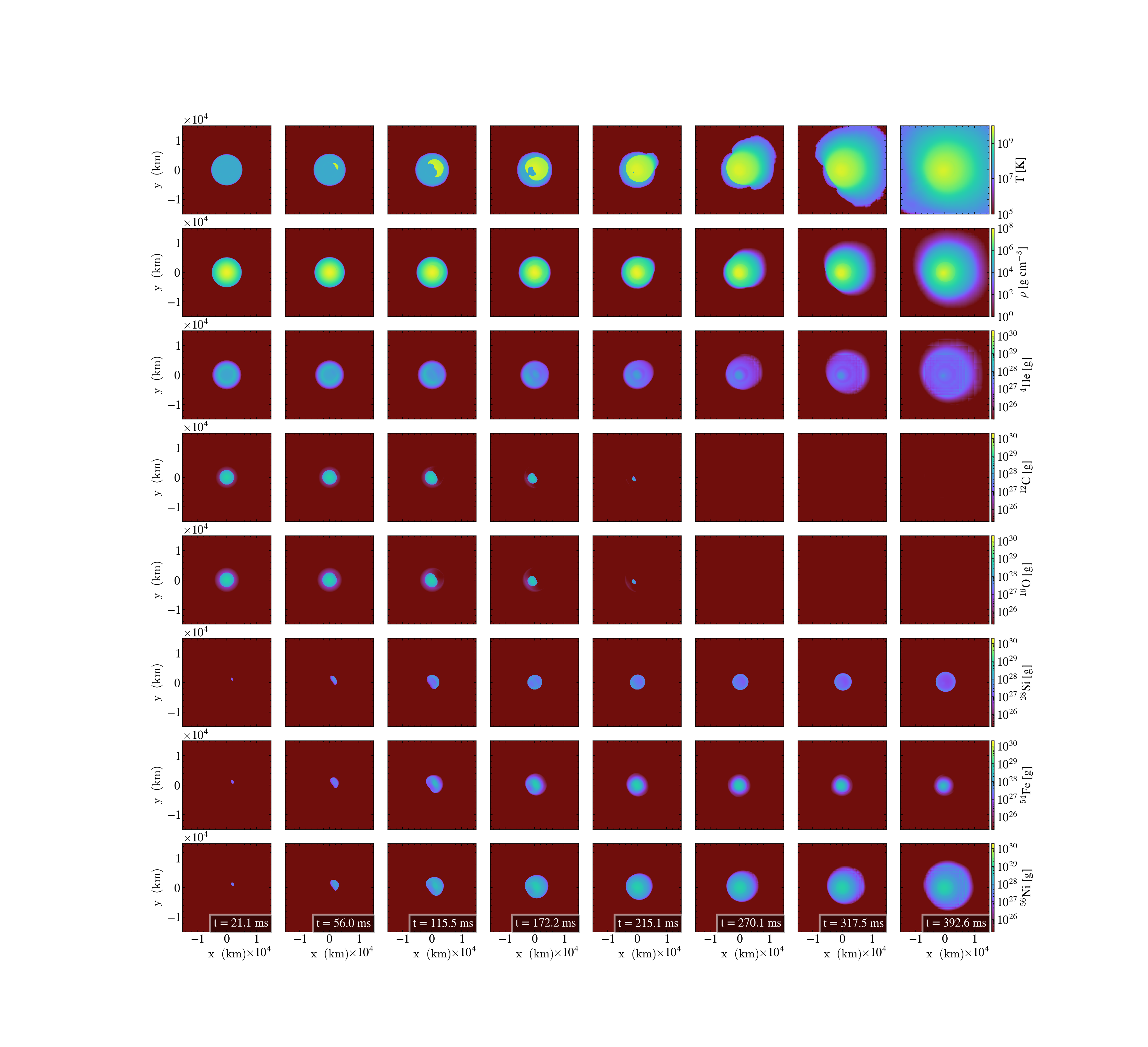}}
  \caption{\textbf{Time sequence of the double detonation.}
  Equatorial $(x\!-\!y)$ slices of temperature and density (top rows), followed by $z$–integrated species maps (He, C, O, Si, Fe, Ni), at
  $t=\{21.1,\,56.0,\,115.5,\,172.2,\,215.1,\,270.1,\,317.5,\,393.6\}$~ms. The helium–shell detonation wraps around the star while an inward shock compresses the CO core and triggers secondary ignition. He/C/O are depleted as Si/Fe/Ni are produced. By $\sim\!0.5$~s the detonation phase is complete and the ejecta accelerate outward.}
  \label{fig:time_series}
\end{figure*}

We begin the explosion simulation from the relaxed mapped model. Our fiducial run yields a clear double–detonation sequence: a helium–shell detonation that sweeps laterally around the star, and an inwardly propagating shock that strengthens by geometric convergence and ignites a secondary CO detonation, unbinding the white dwarf. Figure~\ref{fig:time_series} presents a zoomed time series through this sequence; the full 3D computational domain spans $10^{11}$~cm in each direction, but for clarity we show only the central region. Columns correspond to eight epochs from $t=21.1$ to $393.6$~ms after the temperature perturbation is applied. The top two rows show equatorial $(x\!-\!y)$ slices of temperature and density. Subsequent rows display, for selected species, the $z$–integrated mass (surface mass) distributed across the $(x\!-\!y)$ plane. The temperature panels capture the rapid onset and lateral wrap–around of the He detonation; the density panels show the expected rarefaction behind the front and compression within the core prior to secondary ignition. The composition panels (He, C, O, Si, Fe, Ni) trace the progression from helium combustion in the shell, through intermediate–mass element (IME) production, to iron–group element (IGE) synthesis once the core detonates. As time proceeds, He, C, and O are depleted while Si, Fe, and Ni increase. By the last column, the burning is essentially complete, and the flow is accelerating outward.

We follow the ejecta to $t=10$~s, by which time the expansion is effectively homologous. The velocity–mass relation is monotonic with maximum speeds of $\simeq 2.2\times10^4~\mathrm{km\,s^{-1}}$, consistent with normal SNe~Ia. Any modest asymmetry introduced by the off–axis temperature perturbation is largely erased by the rapid wrap–around of the shell detonation and is not expected to leave observable signatures.

Integrated elemental yields from tracer post–processing are listed in Table~\ref{tab:elements}. The model synthesizes $M(^{56}\mathrm{Ni})\simeq 0.64\,\mathrm{M_\odot}$, indicative of a normal–luminosity SN~Ia. The intermediate–mass element yield is $M_{\rm IME}\simeq 0.41\,\mathrm{M_\odot}$, while the total iron–group mass is $\simeq 0.66\,\mathrm{M_\odot}$. While the progenitor accumulated a substantial helium envelope during its growth phase ($\simeq0.4M_\odot$, as shown in Figure~\ref{fig:wd_prof}), the final ejecta consists of only $M(\mathrm{He})\simeq 6.7\times10^{-3}\,\mathrm{M_\odot}$ unburned He. Residual CO is represented by $M(\mathrm{O})\simeq 9.1\times10^{-4}\,\mathrm{M_\odot}$ and $M(\mathrm{C})\simeq 1.9\times10^{-6}\,\mathrm{M_\odot}$. Trace shell–ash species (e.g., Ti, Cr, Mn, V, K, Cl) are present at the $10^{-6}$–$10^{-2}\,\mathrm{M_\odot}$ level.

\begin{table*}
\centering
\begin{tabular}{cc}
\hline
Atom & Mass ($M_{\odot}$) \\
\hline
$^{28}\text{Ni}$ & $6.4\times10^{-1}$ \\
$^{27}\text{Co}$ & $1.7\times10^{-4}$ \\
$^{26}\text{Fe}$ & $1.8\times10^{-2}$ \\
$^{25}\text{Mn}$ & $4.4\times10^{-5}$ \\
$^{24}\text{Cr}$ & $1.1\times10^{-2}$ \\
$^{23}\text{V}$  & $3.6\times10^{-6}$ \\
$^{22}\text{Ti}$ & $2.2\times10^{-4}$ \\
$^{20}\text{Ca}$ & $4.5\times10^{-2}$ \\
$^{19}\text{K}$  & $2.6\times10^{-6}$ \\
$^{18}\text{Ar}$ & $2.7\times10^{-2}$ \\
$^{17}\text{Cl}$ & $5.2\times10^{-6}$ \\
$^{16}\text{S}$  & $1.2\times10^{-1}$ \\
$^{15}\text{P}$  & $4.0\times10^{-5}$ \\
$^{14}\text{Si}$ & $2.2\times10^{-1}$ \\
$^{12}\text{Mg}$ & $1.8\times10^{-5}$ \\
$^{8}\text{O}$   & $9.1\times10^{-4}$ \\
$^{6}\text{C}$   & $1.9\times10^{-6}$ \\
$^{2}\text{He}$  & $6.7\times10^{-3}$ \\
\hline
\end{tabular}
% \caption{\textbf{Elemental yields (high $Z$ to low $Z$).}
% Tracer–post–processed ejecta masses (in $M_\odot$) for a representative subset of species, ordered by atomic number.}
\caption{\textbf{Elemental yields (high $Z$ to low $Z$).}
Integrated ejecta masses (in $M_\odot$) for selected species, ordered by atomic number $Z$ from highest to lowest. Values are derived from tracer post–processing with a 127–isotope network at the epoch when the ejecta are effectively homologous. Only a representative subset is shown.}
\label{tab:elements}
\end{table*}

\begin{figure*}
  \centering
  \includegraphics[width=0.85\textwidth]{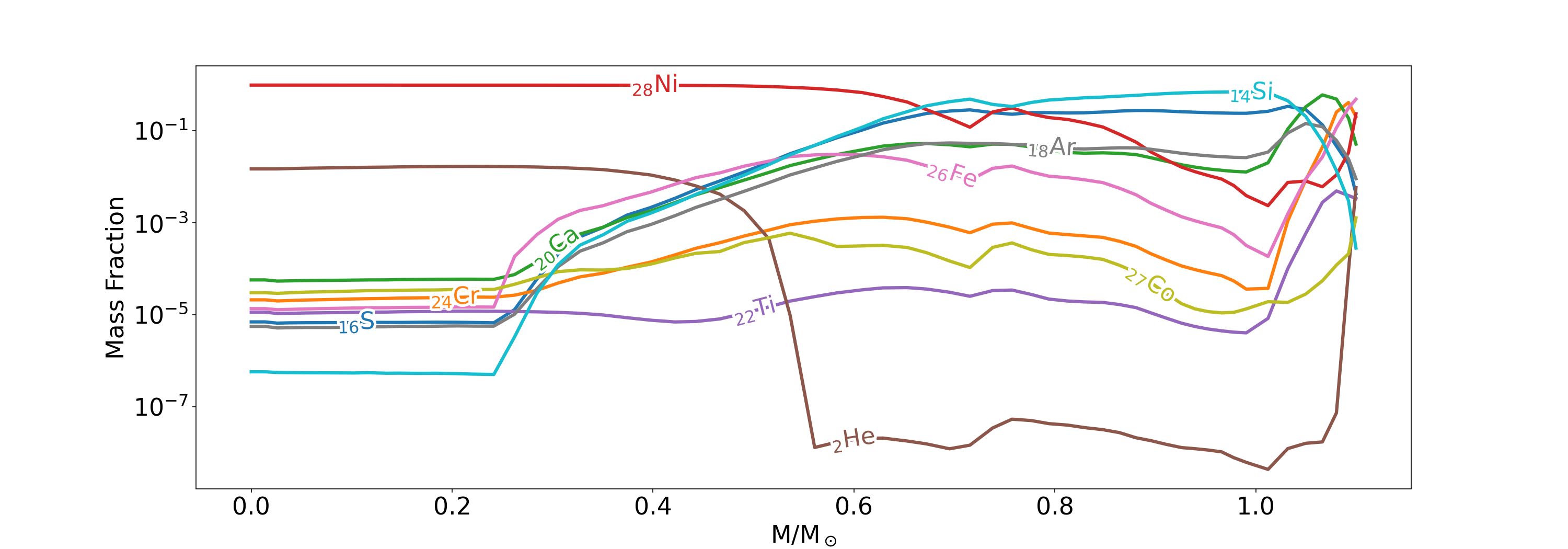}
  \caption{\textbf{Abundance stratification in mass coordinates.}
  Angle–averaged mass fractions binned by the tracers' initial enclosed–mass coordinate $M/M_\odot$, evaluated once the ejecta are homologous. The inner ejecta are dominated by nickel, followed by an IME mantle (Si, S, Ar, Ca), and a thin outer layer of shell ashes.}
  \label{fig:massfrac}
\end{figure*}

The final abundance stratification is shown in Figure~\ref{fig:massfrac}, which plots angle–averaged mass fractions versus the initial enclosed–mass coordinate $M/M_\odot$ associated with each tracer. The inner $\sim$half of the ejecta are nickel–dominated (mass fraction $\approx 1$), transitioning outward to an IME mantle (Si, S, Ar, Ca) and then to a thin skin of shell ashes at the highest mass coordinates. The rise of IMEs beyond $M\!\approx\!0.4$–$0.5\,\mathrm{M_\odot}$ and enhanced Ca/Ar in the exterior are hallmarks of sub–Chandrasekhar double detonations and naturally map to line–forming regions at intermediate and high velocities.

Finally, we calculate the abundance–versus–mass structure to velocity space using the angle–averaged homologous field at the final snapshot. The nickel–dominated core occupies the lowest velocities, $v\!\lesssim\!(7$–$8)\times10^{3}\,\mathrm{km\,s^{-1}}$. The IME mantle (Si, S, Ar, Ca) spans intermediate velocities of order $(8$–$16)\times10^{3}\,\mathrm{km\,s^{-1}}$, naturally encompassing the range where strong Si\,\textsc{ii} and S\,\textsc{ii} features typically form. The He–shell ashes are confined to the outermost layers at $v\!\gtrsim\!(1.6$–$2.2)\times10^{4}\,\mathrm{km\,s^{-1}}$, consistent with the highest ejecta speeds measured in the model. These ranges vary only weakly with viewing angle (a few percent) because the shell detonation wraps around the star rapidly, leaving a largely spherical, layered structure by the time the flow becomes homologous.
\\
\\

\section{Discussion}
\label{sec:discussion}

Our simulation produces a clear double detonation from a single-degenerate, helium-accumulating CO white dwarf, that built up a thick helium envelope by prolonged, quiescent accretion. The layered outcome—an inner ${}^{56}$Ni–dominated core ($\simeq 0.64\,\mathrm{M_\odot}$), an IME mantle ($M_{\rm IME}\simeq 0.41\,\mathrm{M_\odot}$), and high–velocity shell ashes—together with maximum ejecta speeds of $\sim 2.2\times10^4~\mathrm{km\,s^{-1}}$, is characteristic of normal–luminosity SNe~Ia from sub–Chandrasekhar explosions. In this framework, several observational facets emerge naturally.

First, the abundance layering and velocity partition suggest standard photospheric–phase line formation: Si\,\textsc{ii} and S\,\textsc{ii} at $\sim (8$–$16)\times10^3~\mathrm{km\,s^{-1}}$, Ca\,\textsc{ii} features (including potential high–velocity components) at the outer edge of the IME layer and into the shell ashes, and rapid color evolution governed by the ${}^{56}$Ni distribution. The small residual C/O masses ($M(\mathrm{C})\!\sim\!2\times10^{-6}\,\mathrm{M_\odot}$; $M(\mathrm{O})\!\sim\!9\times10^{-4}\,\mathrm{M_\odot}$) imply weak or absent early C\,\textsc{ii} signatures, while the non-negligible Ca and trace Ti ($M(\mathrm{Ca})\!\simeq\!0.045\,\mathrm{M_\odot}$; $M(\mathrm{Ti})\!\simeq\!2.2\times10^{-4}\,\mathrm{M_\odot}$) point to possible high-velocity Ca\,\textsc{ii} and mild early UV/blue suppression. These tendencies align qualitatively with early-excess SNe~Ia where a shallow outer ${}^{56}$Ni component or shell ashes have been invoked, yet they also highlight the degeneracy of interpretations (e.g., the diversity seen in SN~2023bee and transitional events like SN~2022xkq). A decisive test will come from radiative-transfer comparisons using our fully stratified ejecta (density+composition versus velocity) to predict early flux excesses, color evolution, and high-velocity features (HVFs).

Independent, remnant–phase support for this layered structure has emerged from integral–field spectroscopy of the young LMC remnant SNR~0509–67.5. High–resolution MUSE data reveal a \emph{double–shell} morphology in highly ionized calcium ([Ca\,XV]) and a single sulphur shell in the reverse–shocked ejecta. The outer Ca shell is naturally interpreted as the product of the helium–layer detonation, while the inner Ca shell arises from incomplete burning in the CO–core detonation—precisely the stratification expected from sub–Chandrasekhar double detonations. This “double–shell Ca” fingerprint provides spatially resolved, remnant–phase evidence that the double–detonation mechanism operates in nature \citep{Das2025,Soker2025double–shell-Ca}.
Second, the total ${}^{56}$Ni mass is adequate for normal peak luminosities without requiring near-Chandrasekhar densities or strong electron captures. The comparatively low stable iron–peak content ($M(\mathrm{Fe})\simeq 0.018\,\mathrm{M_\odot}$; trace Mn/Cr) is consistent with sub–Chandrasekhar densities at burning and should imprint itself in nebular ratios (e.g., [Fe\,\textsc{iii}]/[Fe\,\textsc{ii}], [Ni\,\textsc{ii}]/[Fe\,\textsc{ii}]) at $\gtrsim 200$~days. Because the small off–axis temperature perturbation is rapidly overtaken by the wrapping He detonation, global asymmetries are muted at homology and are unlikely to leave observable signatures or strong continuum polarization.

Third, in the broader progenitor context, this work strengthens the case that \emph{quiescent} single–degenerate helium accretors can reach double–detonation conditions, even at sub-Chandrasekhar masses. In the 1D evolutionary tracks we build upon, specific combinations of WD mass, core temperature, and $\dot{M}\!\gtrsim\!10^{-8}\,\mathrm{M_\odot\,yr^{-1}}$ allow the system to avoid discrete He–nova ejections, steadily amassing a shell until runaway. Our mapped profiles thus provide physically motivated seeds for the hydrodynamic phase, as opposed to ad hoc shell prescriptions. From a demographics perspective, such systems would naturally populate intermediate delay times and evade many SD observational constraints (no H lines, faint/compact donors, low CSM signatures), complementing double–degenerate channels and potentially helping to reconcile the observed diversity of early light curves and spectra.

Potential helium donors in the single-degenerate path include (i) compact, non–degenerate He stars (e.g., post-interaction sdB/sdO–like objects) transferring through Roche-lobe overflow; (ii) stripped subgiants with He–rich envelopes; (iii) H-nova producing systems where the average cyclic accretion rate is high enough to leave a He-rich residue that accumulates over hundreds (or thousands) of eruptions; and (iv) H–rich donors that, under steady H burning, quietly build a He layer on the WD over long intervals. \cite{hillman2025} showed that at $\dot{M}\!\approx\!10^{-8}\,\mathrm{M_\odot\,yr^{-1}}$ a CO WD can, under specific conditions, continue helium accumulation without triggering recurrent He novae, naturally reaching the pre–detonation structures we employ here. Observationally, such donors can be UV–bright yet optically modest; combined with the lack of strong winds, they are consistent with the non–detections of luminous companions and tight radio/X–ray limits in nearby SNe~Ia. Future constraints may come from deep pre–explosion UV imaging in star–forming hosts, as well as post–explosion searches for faint survivors.

\subsection{Caveats}
Naturally, our models hold various caveats.
While our network captures the dominant energy release, detailed isotopic ratios (e.g., stable Ni and Mn/Cr) are best constrained through tracer post–processing and full radiative–transfer modeling for direct comparison with nebular spectra. Resolution and perturbation–placement indicate robustness of the outcome (He detonation followed by core ignition), but a broader parameter study—including shell mass, WD mass, and accretion history—will quantify the breadth of normal versus sub–luminous outcomes and the strength of early bumps and HVFs. Finally, connecting these models to rates requires coupling binary–population synthesis for helium donors with retention efficiencies calibrated to long–term He–nova evolution. To complete the picture, the ``undetermined transient" branch identified by \citet{hillman2025} under comparable accretion conditions should also be subjected to the same multi-D explosion and radiative-transfer analysis, to determine whether it leads to faint thermonuclear transients or non-explosive outcomes.

\subsubsection{IME/IGE in the outer ejecta}
\label{sec:caveats}
A known vulnerability of thick-shell double detonations is the chemical makeup of
the outer $v\!\gtrsim\!1.5\times 10^4~\mathrm{km\,s^{-1}}$ layers. He detonations at
$\rho\!\sim\!10^5$--$10^6~\mathrm{g\,cm^{-3}}$ can synthesize significant Ca and, depending
on pollution, Ti/Cr, while leaving some residual He 
\citep{WoosleyWeaver1994,ShenMoore2014}. These species shape early UV/blue opacity and
can drive prominent high-velocity Ca\,\textsc{ii} features. Grids and case studies indicate
that more massive shell or shells with low C/N pollution tend to produce stronger UV
suppression and redder $g-r$ at early times, with SN~2018byg providing an extreme instance
\citep{Polin2019,De2019}.

To make these risks explicit, we tabulate (i) $M(\mathrm{Ca}), M(\mathrm{Ti}), M(\mathrm{Cr})$
above $1.5\times 10^4~\mathrm{km\,s^{-1}}$, (ii) the residual He mass in the same velocity
layer, and (iii) the angle-averaged velocity distributions of these species (Table~\ref{tab:elements}
and Fig.~\ref{fig:massfrac}). While radiative transfer is deferred, these numbers can be read
against published thresholds for strong line blanketing and HVFs. We also note that modest
shell pollution by C/O or $^{14}$N reduces Ti/Cr yields and can lower He\,\textsc{i}
detectability by altering ionization/excitation conditions \citep{Boyle2017,Jiang2017,Magee2022}.

\section{Conclusions and Summary}
\label{sec:conclusions}

We have modeled a single–degenerate, sub–Chandrasekhar double detonation by mapping time–evolved helium–accretion profiles into multi–D hydrodynamics and following the ejecta to homology. Our main conclusions are:

(1) \emph{Clear double–detonation sequence.} A laterally propagating helium–shell detonation launches an inward shock that converges and ignites the CO core, unbinding the star. Burning is essentially complete by $\sim 0.5$~s, and by $t=10$~s the flow is homologous.

(2) \emph{Yields consistent with normal SNe~Ia.} The explosion synthesizes $M(^{56}\mathrm{Ni})\simeq 0.64\,\mathrm{M_\odot}$ and $M_{\rm IME}\simeq 0.41\,\mathrm{M_\odot}$, with a total iron–group mass of $\simeq 0.66\,\mathrm{M_\odot}$. Residual unburned masses are small ($M(\mathrm{He})\simeq 6.7\times 10^{-3}\,\mathrm{M_\odot}$, $M(\mathrm{O})\simeq 9.1\times 10^{-4}\,\mathrm{M_\odot}$, $M(\mathrm{C})\simeq 1.9\times 10^{-6}\,\mathrm{M_\odot}$), and maximum ejecta speeds reach $\simeq 2.2\times 10^4~\mathrm{km\,s^{-1}}$.

(3) \emph{Layered ejecta and kinematic partition.} The nickel–dominated core occupies the lowest velocities, an IME mantle (Si, S, Ar, Ca) spans $(8$–$16)\times10^3~\mathrm{km\,s^{-1}}$, and He–shell ashes reside at $\gtrsim(1.6$–$2.2)\times10^4~\mathrm{km\,s^{-1}}$. The rapid wrap–around of the shell detonation erases most large–scale asymmetries, so strong viewing–angle effects are not expected.

(4) \emph{Early–time spectral tendencies.} The enhanced Ca and trace Ti in the outer layers imply strong Ca\,\textsc{ii} H\&K (and possible HVFs) and mild UV/blue suppression from line blanketing at early times, fading as the photosphere recedes beneath the shell–ash layer. These signatures, together with the stratified IME/IGE structure, provide testable predictions for light curves and spectra.

(5) \emph{Progenitor implications.} Prolonged, quiescent helium accretion at $\dot M\lesssim 10^{-8}\,\mathrm{M_\odot\,yr^{-1}}$ can build the pre–detonation shells required for double detonations, consistent with compact He–star donors or He-rich channels discussed in prior evolutionary work. Such systems naturally avoid many classic SD constraints (no H lines, weak CSM) while contributing to the diversity of early–time behaviors.

Beyond these findings, our model aligns qualitatively with remnant–phase evidence for layered double detonations (e.g., a double–shell calcium morphology), encouraging targeted radiative–transfer comparisons. Immediate next steps include (i) synthetic light curves and spectra from the stratified ejecta for direct confrontation with high–cadence data; (ii) a parameter survey in WD mass, shell mass, and accretion history to chart ${}^{56}$Ni and IME yields, early bumps, and HVFs; and (iii) coupling to binary–population synthesis for helium donors, with retention efficiencies informed by long–baseline helium–nova evolution, to assess rates and delay–time distributions. Together, these efforts will clarify how frequently \emph{quiescent} helium accretors produce normal SNe~Ia and what observational features most clearly distinguish this channel.

\section*{Acknowledgments}
We acknowledge support for this project from the European Union's Horizon 2020 research and innovation program under grant agreement No 865932-ERC-SNeX.
This work was supported by the Azrieli College of Engineering – Jerusalem Research Fund.

\bibliography{rfrncs}{}
\bibliographystyle{aasjournal}

\end{document}